\documentclass[aps,prb,twocolumn,superscriptaddress,showpacs]{revtex4}
\usepackage{bm}
\usepackage{amsmath}
\usepackage{amssymb}
\usepackage{graphicx}
\usepackage{color}
\input{epsf}

\begin{document}

\date{\today}
\title{Entanglement between charge qubit states and coherent
states of nanomechanical resonator generated by AC Josephson
effect}

\author{O.M.Bahrova}
\affiliation{B. Verkin Institute for Low Temperature Physics and
Engineering of the National Academy of Sciences of Ukraine, 47
Prospekt Nauky, Kharkiv 61103, Ukraine}

\author{L.Y.Gorelik}
\affiliation{Department of Physics, Chalmers University of
Technology, SE-412 96 G{\"o}teborg, Sweden}

\author{S.I.Kulinich}
\affiliation{B. Verkin Institute for Low Temperature Physics and
Engineering of the National Academy of Sciences of Ukraine, 47
Prospekt Nauky, Kharkiv 61103, Ukraine}

\date{\today}

\begin{abstract}
We considered a nanoelectromechanical system consisting of a movable
Cooper-pair box qubit which is subject to an electrostatic field,
and coupled to the two bulk superconductors via tunneling processes.
We suggest that qubit dynamics is related to the one of a
quantum oscillator and demonstrate that a bias voltage applied
between superconductors generates states represented by the
entanglement of qubit states and coherent states of the oscillator
if certain resonant conditions are fulfilled.   It is shown that a
structure of this entanglement may be controlled by the bias voltage
in a way that gives rise to the entanglement incorporating so-called
cat-states - the superposition of coherent states.  We characterize
the formation and development of such states analyzing the entropy
of entanglement and corresponding Wigner function.  The
experimentally feasible detection of the effect by measuring the
average current is also considered.

\end{abstract}
\maketitle

\section*{Introduction}
Electro-mechanical phenomena on the nanometer scale attract
significant attention during the last two decades. \cite{Roukes}
Recent advantages in nanotechnologies acquire a promising platform 
for studying the fundamental phenomena generated by the interplay between quasi-classical and
pure quantum subsystems. A charge qubit formed by a tiny
superconducting island (Cooper-pair box (CPB)) whose basis states
are charge states (e.g. states which represent the presence or
absence of excess Cooper pairs on the island), is one of a large
group of pure quantum systems. \cite{Pashkin} At the same time
modern nanomechanical resonators which dynamics according to
Ehrenfest theorem to great extent is described by classical
equations, are ideal representatives of quasiclassical subsystem
\cite{Schmid}. Systems, which dynamics is determined by the mutual influence between a superconducting qubit and a nanomechanical resonator, are a subject of cutting-edge research in quantum physics, especially, in quantum communication, see, e.g., Refs.~\cite{Cleland1,Cleland2,Chu,Cleland3,LaH,Tian}

There are two main questions that arise related to an
interplay between quasi-classical dynamics of the mechanical
resonator and quantum dynamics of the charge qubit. The first one
is: how quasi-classical motion may affect pure quantum
phenomena? Considering this question, it was shown that the
superconducting current between two remote superconductors can be
established by mechanical transportation of Cooper pairs performed
by an oscillating CPB. \cite{Gorelik2} Even more, it was
demonstrated that such transportation may generate correlations
between the phases of space-separated superconductors.
\cite{Isacsson} Another question is how coherent Josephson dynamics
of a charge qubit will affect the dynamics of the quasi-classical
resonator, in particular, whether or not the quantum
entanglement between a superconducting qubit and mechanical vibrations can be
achieved?  Recently it was demonstrated that individual phonons can
be controlled and detected by a superconducting qubit enabling
coherent generation and registration of quantum superposition of
zero and one-phonon Fock states \cite{Cleland1,Cleland2}.  At the
same time nanomechanical resonators provide the possibility to store
quantum information in the complex multi-phonon coherent states.
Such states, in contrast to single-phonon states, where mechanical
losses irreversibly delete the quantum information, allow their
detection and correction \cite{Girvin, Girvin1}. Motivated by such a
challenge, in this paper, we demonstrate the possibility to generate
quantum entanglement between the charge qubit states and mechanical coherent ones in
a particular nanoelectromechanical system (NEMS) where mechanical
vibrations are highly affected due to the weak coupling with movable a
Cooper-pair box.

\section*{Model and Hamiltonian}

\begin{figure}
\includegraphics[width=0.85\columnwidth]{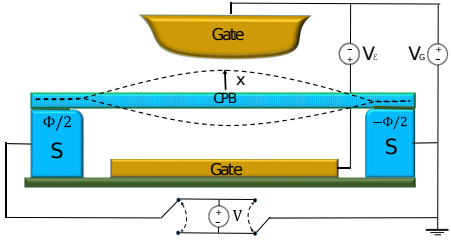}
\caption{Schematic illustration of the NEMS under consideration.
The superconducting nanowire, treated as a charge qubit, is tunnel
coupled to two bulk superconductors (S) with the superconducting phase
difference $\Phi$ and capacitively coupled to the two gate electrodes.
The bending oscillations in the $x$ direction are described by the
harmonic oscillator.}\label{model}
\end{figure}

Schematic representation of the NEMS prototype considered in this
article is presented in Fig.~\ref {model}. It consists of the
superconducting nanowire (SCNW),\cite{NW1,NW2}, which is suspended
between two bulk superconductors and is capacitively coupled to the
two side gate electrodes. In this paper, we will consider the case
when SCNW represents a superconducting island that can be treated as
a charge qubit (Cooper-pair box) whose basis states are charge
states - states which represent the presence or absence of excess
Cooper pairs on the island. Below we will refer to these states as
charge and neutral states correspondingly. As this takes place, the
gate voltage $V_{G}$ and the voltage applied between the gates
$V_{\mathcal{E}}$ are chosen in a way that the difference in the
electrostatic energies of the charged and neutral states equals to
zero at the straight configuration of the nanowire, while nanowire
bending removes this degeneracy. We also reduce the bending dynamics
of the SCNW to the dynamics of the fundamental flexural mode
described by the harmonic oscillator.
Joint  Cooper pairs dynamics and mechanical one of this system
is described by the Hamiltonian which can be presented in the form,
\begin{eqnarray}\label{1}
&&H=H_q+H_m+H_{int}, \\
&&H_q=-E_J\sigma_1\cos\Phi ,\quad
H_m=\frac{\hbar\omega}{2}\left(\hat x^2+\hat p^2\right),\nonumber\\
&&H_{int}= \varepsilon\hat x\sigma_3.\nonumber
\end{eqnarray}
Here Hamiltonian $H_q$ represents Josephson coupling between CPB and
bulk superconductors. The constant $E_J$ is the Josephson coupling
energy (in this paper we will consider only the case of symmetric
coupling), $\Phi=\Phi (t)$ is the superconducting phase difference
between electrodes, $\sigma_i (i=1,2,3)$ are the Pauli matrices
acting in the qubit Hilbert space in a basis where vectors $(1,0)^T$
and $(0,1)^T$ represent charged and neutral states, respectively.
Hamiltonian $H_m$ in Eq.~(\ref{1}) represents dynamics of the
fundamental bending mode described by the harmonic oscillator with
frequency $\omega$ (here momentum and coordinate operators, $\hat p$
and $\hat x$, are normalized on the amplitude of zero-point
oscillations $x_0=\sqrt{\hbar/M\omega}$, $M$ is an effective mass,
$[\hat x,\hat p]=\imath$). The third term, $H_{int}$, describes an
electromechanical coupling between the charge qubit and the mechanical
oscillator induced by the electrostatic force acting on the charged
state of the qubit, $\varepsilon =e\mathcal E x_0$. In the last
equality, $\mathcal E$ is an effective electrostatic field that is
controlled by the difference of the applied voltages $V_G$ and
$V_\varepsilon$. Below we will assume $\varepsilon\ll\hbar \omega,
E_J$ that corresponds to the typical experimental situation.\cite{Cleland1,Tian,A-A}

The states of the system described by the Hamiltonian, Eq.~(\ref{1}),
are a superposition of direct products  of qubit states, $\mathbf
e^\pm_i$, and  eigenstates  of the oscillator $\vert n\rangle$ (here
and below $\mathbf e^\kappa_i$ denotes the eigenvectors of the Pauli
matrices $\sigma_i$ with eigenvalues $\kappa=\pm 1$).

If $\varepsilon=0$, the interaction between the qubit and the
mechanical subsystem is switched off and stationary states of the
Hamiltonian, Eq.~(\ref{1}), are pure states (the entropy of
entanglement is an integral of motion, i.e. if the system is
initially, in a pure state, it will be in a pure state at any moment
of time). Synchronous switching on the electrical field $\mathcal E$ and
the bias voltage between superconducting leads ($\dot \Phi(t)=
2eV/\hbar$) results in the evolution of such pure states in the
states represented by entanglement between the qubit and
oscillator states.

\section*{Time evolution }

To carry out an analysis of this evolution, we introduce the
dimensionless time and energies, $\omega t \rightarrow t,
E_J/\hbar\omega \rightarrow E_J, \varepsilon/\hbar\omega
\rightarrow\varepsilon$ and assume that at the moment of switching
on the interaction between the subsystems ($ t = 0 $), the
difference between the superconducting phases is $\Phi = \Phi_0$ and
the system has been in a pure state,
\begin{equation}\label{in}
\vert \Psi(0)\rangle = \mathbf e_{in} \otimes \vert
0\rangle.
\end{equation}
At $t>0$ according to Josephson relation
$\Phi(t)=2eVt/\hbar\omega+\Phi_0$. The Hamiltonian, Eq.~(\ref{1}),
and, as a consequence, the time evolution operator $\hat U(t,t')$,
which is defining evolution of the arbitrarily initial state, has the
properties:
\begin{equation}\label{1a}
\hat{H}(t+T_V)=\hat H(t), \quad \hat U(t,t')=\hat U(t+T_V,t'+T_V),
\end{equation}
where $T_V=2\pi/\Omega_V=\pi\hbar\omega/e|V|$. To analyze the
evolution operator, one can use the interaction picture taking
\begin{equation}\label{6}
\hat U(t,t')= \hat{\mathcal U}_\eta (t) \hat{\mathcal U}_\eta
(t,t')\hat{\mathcal U}^\dag_\eta(t'),
\end{equation}
where
\begin{equation}\label{7}
\hat{\mathcal U}_\eta(t)=\exp\left[ \frac{\imath
E_J}{\Omega_V}\sigma_1 \sin\left(\Omega_V
t+\eta\Phi_0\right)-\imath a^\dag a t\right].
\end{equation}
The parameter $\eta=\text{sgn}\left(V/|V|\right)=\pm$
characterizes the direction of the bias voltage drop. The operator
$\hat{\mathcal U}_\eta (t,t')$ obeys the following equation:
\begin{eqnarray}\label{9}
\imath \frac{\partial \hat{\mathcal U}_\eta(t,t')}{\partial t}=
\hat{\mathcal H}_\eta(t)\hat{\mathcal U}_\eta(t,t'),\nonumber \\
\hat{\mathcal H}_\eta(t)= \varepsilon \hat x(t)\sigma_3(t),\quad
\hat{\mathcal U}_\eta(t,t)=\hat I.
\end{eqnarray}
Here
\begin{eqnarray}
&&\hat x(t) = \frac{1}{\sqrt 2}(\hat a \text{e}^{-\imath t}+\hat a^\dag
\mathrm{e}^{\imath t}),\nonumber\\
&&\sigma_3(t)=\sigma_3\cos\left(\frac{E_J}{\Omega_V}
\sin(\Omega_V t+\eta\Phi_0)\right)- \nonumber\\
&&\hspace{1cm}-\sigma_2\sin\left(\frac{E_J}{\Omega_V}
\sin(\Omega_Vt+\eta\Phi_0)\right).
\end{eqnarray}

If the frequencies $\omega$ and $\Omega_{V}$ are incommensurable, the
operator $\hat{\mathcal H}_\eta(t)$ is a quasiperiodic function of
time. In such a case one can expect that the mechanical subsystem,
being initially in the ground state, does not significantly deviate
from this state in the process of evolution. However, a rigorous
consideration of this case requires independent research and will be
done elsewhere. In this paper, we will consider the resonant case
when $\Omega_V=\omega$ and will assume that $\varepsilon \ll 1$. The
first condition stipulates the following properties of the evolution
operator,

\begin{equation}\label{N}
\hat {\mathcal U}_\eta(2\pi N, 2\pi N') =\left(\hat{\mathcal
U}_\eta(2\pi,0)\right)^{N-N'},
\end{equation}
where $N, N'$ are the  natural numbers. The second assumption
allows us to make the following substitution in a leading
approximation regarding small $\varepsilon$,
\begin{equation} \hat{\mathcal U}_\eta(t,t')
= \hat{\mathcal U}_\eta (2\pi N,2\pi N'),
\end{equation}
where $N (N')=[t(t')/2\pi] ([x]$ is an integer part of $x$), and
obtain an expression for $\hat{\mathcal U}_\eta(2\pi,0)$ which can
be written as,
\begin{eqnarray}\label{Evop}
&&\hat{\mathcal U}_\eta(2\pi,0)= \exp\left[\imath\tilde\varepsilon
\sigma_2\hat p(\eta\Phi_0)+\varepsilon^2{\cal O}(\hat I)\right], \nonumber\\
&&\hat p(\Phi)= \hat p\cos\Phi+\hat x\sin \Phi.
\end{eqnarray}
Here $ \tilde\varepsilon= 2\pi \varepsilon J_1(2E_J)$ and $J_1(x)$
the Bessel function of the first kind. Using the above relations one
can obtain an expression for the evolution operator $\hat U(t,t')$,
which in the main approximation regarding $\varepsilon $ has a form,
\begin{equation}\label{U}
\hat U(t,t')=\hat{\mathcal U}_\eta(t)
\exp\left[\imath\tilde\varepsilon \sigma_2\hat
p(\eta\Phi_0)(t-t')\right]\hat{\mathcal U}^\dag_\eta(t').
\end{equation}

Using Eqs.~(\ref{in}),(\ref{U}) one gets that at the time $t$, with
the accuracy to small parameter $\tilde\varepsilon\ll 1$, the state
of the system $\vert \Psi(t)\rangle$ is given by an expression,
\begin{equation}\label{WF}
\vert \Psi(t)\rangle=\sum\limits_\kappa
A_\kappa^\eta\mathbf{e}^\kappa_2(t,\eta\Phi_0)\otimes\vert -\kappa
\mathrm{z}(t,\eta)/\sqrt 2 \rangle.
\end{equation}
Here
$$\mathbf{e}^\kappa_2(t,\eta\Phi_0)=\mathbf{e}^\kappa_2
\exp\left[\imath E_J\sigma_1\sin(t+\eta\Phi_0)\right]$$
and $\mathbf{e}_2^\kappa=\sigma_1\mathbf{e}_2^{-\kappa}$ are the
eigenvectors of Pauli matrix $\sigma_2$ with eigenvalues $\kappa
=\pm 1, A_\kappa^\eta=\left(\mathbf{e}_2^\kappa
(0,\eta\Phi_0),\mathbf{e}_{in}\right)$. The symbol
$|\alpha\rangle$ (where $\alpha$ is a complex number) denotes the
coherent states of the oscillator, $\hat a|\alpha\rangle=
\alpha|\alpha\rangle$, while a complex function
$\mathrm{z}(t,\eta)$ is defined as
\begin{equation}\label{670}
\mathrm{z}(t,\eta)=\tilde \varepsilon
t\exp\left[-\imath(t+\eta\Phi_0)\right].
\end{equation}

It should be stressed that Eq.~(\ref{WF}) is valid only for
restricted time interval $t\leq \tilde\varepsilon^{-2}$. Time $t$ should be
also shorter than any dephasing and relaxation times. From
Eq.~(\ref{WF}) one can see that initially pure state
$\vert \Psi(t=0)\rangle=\mathbf{e}_{in}\otimes |0\rangle$
evolves into the state represented by the entanglement between the
two qubit states and two coherent states of the mechanical
resonator. Moreover, the details of this entanglement depend on
switching time (parameter $\Phi_0$) and direction of the bias
voltage (parameter $\eta$). These circumstances allow one to
manipulate the described above entanglement by switching the bias
voltage direction.

\section*{Generation of ``cat-states''}

To demonstrate the effect of the entanglement between the charge qubit and mechanical vibrations that comprehends the formation of so-called Schr\"{o}dinger-cat states of nanomechanical resonator, we consider the following time protocol
for $V(t)$:
$$2eV(t)=-\hbar\omega\theta(t)\left[1-2\theta(t-t_s)\right].$$
Namely, during the time interval $0<t<t_s$ the bias voltage
$V(t)=-\hbar\omega/2e$ and then it switches its sign. Using
Eqs.~(\ref{6}), (\ref{N}), (\ref{Evop}), one gets that at $t>t_s$,
\begin{eqnarray}\label{Sw1}
&&\hat U(t,0)=\hat{\mathcal {U}}_+(t)
\text{e}^{\imath\sigma_2\tilde\varepsilon (t-t_s)\hat
p(\Phi_0)}\hat S \text{e}^{\imath\sigma_2\tilde\varepsilon t_s
\hat p(-\Phi_0)}\hat{\mathcal{U}}_-(0), \nonumber\\
&& \hat S=\hat{\mathcal{U}}_+^\dag (t_s)\hat{\mathcal{U}}_-(t_s)
\equiv\rho(t_s,\Phi_0)+\imath\tau(t_s,\Phi_0)\sigma_1,\\
&&\rho(t_s,\Phi_0)=\cos\left( 2E_J\cos
t_s\sin\Phi_0\right),\nonumber\\
&&\tau(t_s,\Phi_0)=-\sin\left(2E_J\cos t_s\sin \Phi_0\right).
\nonumber
\end{eqnarray}
As a result, the state of the system after changing the direction of
the bias voltage has a form:
\begin{eqnarray}\label{WF1}
&&\hspace{-1.1cm}|\Psi(t)\rangle=
\sum\limits_\kappa\mathbf{e}^\kappa_2(t,\Phi_0)
\otimes\nonumber\\
&&\otimes\left(\rho A_\kappa^-|-\kappa \mathrm{z}_+/\sqrt
2\rangle+\imath\tau A_{-\kappa}^-|\kappa \mathrm{z}_-/\sqrt
2\rangle\right),
\end{eqnarray}
where $\mathrm{z}_\pm=\mathrm{z}_1\pm\mathrm{z}_2$ and
\begin{equation}\label{23}
\mathrm{z}_1=\text{e}^{-\imath(t-\Phi_0)}\tilde\varepsilon t_s,
\mathrm{z}_2=\text{e}^{-\imath(t+\Phi_0)}\tilde\varepsilon(t-t_s)
\end{equation}
(see Fig.~2). It can be seen from this equation that the state of the
system is represented by the entanglement of two qubit's state with
two so-called ``cat-states" (superposition of coherent states) whose
structure is controlled by the parameters $E_J\ (\rho )$ and $\Phi_0$. As it
follows from Eqs.~(\ref{WF1}), (\ref{23}), the bias voltage switching
does not affect the dynamics of the system if $\Phi_0= 0,\pi$.
\begin{figure}
\includegraphics[width=.85\columnwidth]{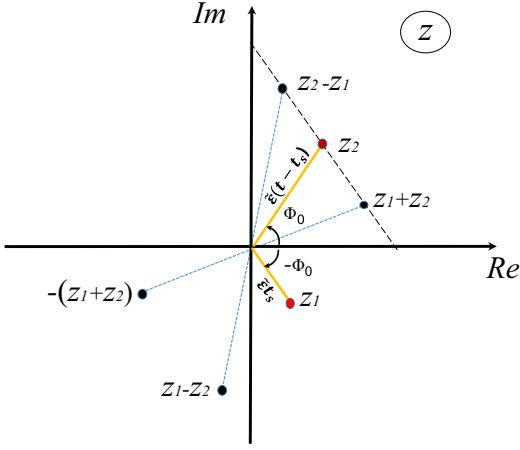}
\caption{Schematic illustration of the positions of the coherent
states described by the complex numbers $\mathrm{z}_{1,2}$ and their
combinations $\mathrm{z}_\pm$ in the complex plane. It denotes the time evolution of the coherent states, on the one hand, and the dependence on the initial phase difference $\Phi_0$, on the other one.
}\label{Fig2}
\end{figure}

Below we will limit ourselves to considering a most interesting,
from our point of view, case when $\Phi_0 = \pi/2$ and put
$\mathbf{e}_{in} = (\mathbf{e}_2^+ + \mathbf{e}^-_2)/\sqrt 2$, that
is, we suppose that immediately before the interaction was switched
on, the qubit was in the eigenstate of the operator $\hat
H_q(t=0-\delta)$. These assumptions lead to the following relations
$A_+^-=A_-^-=\exp(\imath E_J)/\sqrt 2$  in Eq.~(\ref{WF1}). To
characterize the entanglement between the qubit states and  the
states of the mechanical oscillator, we introduce the reduced
density matrices,
$\hat\varrho_{q(m)}(t)=\mathrm{Tr}_{m(q)}\hat\varrho$, where
$\hat\varrho=|\Psi(t)\rangle\langle\Psi(t)|$ is a
complete density matrix of the system and $\mathrm{Tr}_{m(q)}$
denotes the trace over mechanical (qubit) degrees of freedom. Using Eqs.~
(\ref{WF},\ref{WF1}), one can get the following expression for the
$\hat\varrho_q$,
\begin{equation}\label{54}
\hat\varrho_q(t)=\frac{ I+\lambda(t,t_s)\sigma_1}{2},
\end{equation}
where
\begin{eqnarray} \label{DMQ}
&& \lambda(t,t_s)=\exp{\left(-\tilde{\varepsilon}^2t^2\right)},
\hspace{1.0cm} 0<t\le t_s, \\
&& \lambda(t,t_s)=\rho^2\exp{\left[
-\tilde\varepsilon^2(t-2t_s)^2\right]}+\nonumber\\
&&\hspace{1.2cm}+\tau^2\exp{\left( -\tilde{\varepsilon}^2t^2
\right)},\hspace{1.3cm} t>t_s. \label{RHO}
\end{eqnarray}
In deriving this equation, we took into account relation
$\mathbf{e}_2^+\cdot\mathbf{e}_2^-+\mathbf{e}_2^-
\cdot\mathbf{e}_2^+=\sigma_1$. Using Eq.~(\ref{54}) one can
calculate the entropy of entanglement,
\begin{equation}\label{667}
S_{en}(t) \equiv -\operatorname{Tr} \hat\varrho_q(t) \log
\widehat{\varrho}_q(t) =- \operatorname{Tr} \hat\varrho_m(t) \log
\widehat{\varrho}_m(t).
\end{equation}

\begin{figure}
\includegraphics[width=0.85\columnwidth]{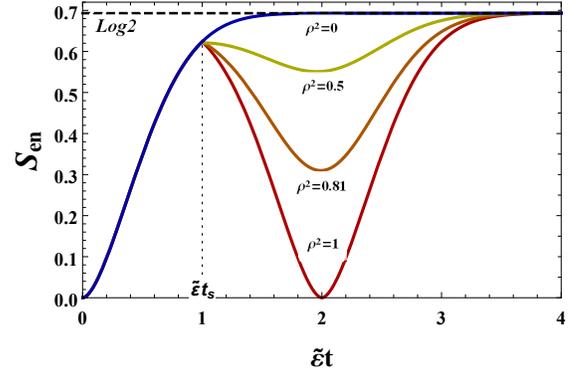}
\caption{The entanglement entropy dependent on time (in units of
$\tilde\varepsilon\omega$), for different values of
$\rho=0,1/\sqrt{2},0.9,1$ (blue, yellow, orange and red curves
online). The thin dotted line indicates the bias voltage switching time. The dashed curve corresponds to the maximal value of the entanglement, log$2$.}
\label{Fig-entropy}
\end{figure}

One can find that $S_{en}(t)$ monotonically increases in time within
intervals $0<t<t_s$ and $2t_s<t<\infty$ saturating to the maximal
value  $S_{en}^{\text{(max)}}=\log 2$ at $t\rightarrow \infty$.
Within interval $t_s<t\leq 2t_s$ the behavior of the entanglement
entropy depends on the relation between $\rho$ and $\tau$. In
particular, for $\rho^2>\tau^2$ the entanglement entropy $S_{en}(t)$
starts to decrease after switching, reaching some minimal value
(equals zero for the $\rho^2=1$) within interval $t_s<t\leq 2t_s$. If
$\rho^2<\tau^2$, the entropy continues to grow just after the switching.
However, its derivative might be also negative within some time
interval whose existence is controlled by the parameters
$\tilde{\varepsilon}t_s$ and $\tau^2/\rho^2$. The plot of
$S_{en}(t)$ for $\tilde{\varepsilon} t_s=1$ and different values of
$\rho$ is presented in Fig.~\ref{Fig-entropy}.

-----------------------------------------------------------------------
\section*{Evolution of mechanical subsystem and average current}

To describe the evolution of the mechanical subsystem, we consider
the reduced density matrix $\widehat{\varrho}_m(t)$. From
Eq.~(\ref{WF1}) one gets that at $t>t_s$:
\begin{eqnarray}\label{DMM}
&&\hat{\varrho}_m(t)=\nonumber\\
&&\frac{1}{2}\sum\limits_\kappa\left[\rho^2|\kappa
\mathrm{z}_+/\sqrt 2\rangle \langle\kappa \mathrm{z}_+/\sqrt
2|+\tau^2|\kappa \mathrm{z}_-/\sqrt 2\rangle\langle\kappa
\mathrm{z}_-/\sqrt 2|- \right.\nonumber\\
&&\left.-\imath \rho\tau \left(|-\kappa \mathrm{z}_+/\sqrt
2\rangle\langle\kappa \mathrm{z}_-/\sqrt 2|-
\text{H.c.}\right)\right].
\end{eqnarray}

To visualize the state of the mechanical subsystem, it is convenient
to use the Wigner function representation for the density matrix
$\hat\varrho_m(t)$,
$$W(x,p,t)=\frac{1}{\pi}\int\varrho_m(x+y,x-y,t)\exp(2\imath py)dy,$$
where $\varrho_m(x,x',t)=\langle x|\hat\varrho_m(t)|x'\rangle $.
Using Eq.~(\ref{DMM}), one gets,
\begin{equation} \label{WiF}
W(x,p,t)= W_t(x\cos t -p\sin t, p\cos t +x\sin t),
\end{equation}
where the function $W_t(x,p)$ is defined according to the relation,
\begin{eqnarray} \label{76}
&&W_t(x,p)=\nonumber\\
&&\frac{1}{2}\sum\limits_\kappa\left[\rho^2
W_0(x,p+\kappa|\mathrm{z}_+|)
+\tau^2W_0(x,p-\kappa|\mathrm{z}_-|)+ \right.\nonumber\\
&&\left. +2\rho\tau \sin\left(2\kappa
Z_-x\right)W_0\left(x,p+\kappa Z_+\right)\right].
\end{eqnarray}
In Eq.~(\ref{76}) $ Z_\pm= \left(
|\mathrm{z}_-|\pm|\mathrm{z}_+|\right)/2$  and
\begin{equation}\label{89}
W_0(x,p)=\frac{1}{\pi}\exp\left[-(x^2+p^2)\right]
\end{equation}
is the Wigner function corresponding to the ground state of the
oscillator. The plot of $W(x,p,t)$ for $t=2\pi N$, $\rho=0,
\rho=1$ and $\rho=\tau=1/\sqrt 2$ at $|\mathrm{z}_+|=3$ and
$|\mathrm{z}_-|=9$ is presented in Figs.~4,5.

From Eq.~(\ref{DMM}),(\ref{76}) one can see that in the case when $ \rho $ is
equal to zero or one (in particular, when $ t_ s = 0$) the Wigner
function is positive and has two maxima, demonstrating the
entanglement between two states of the qubit and two coherent states
(see Fig.~4). In general case $\rho\tau\neq 0$, and the Wigner function
takes both positive and negative values at $t>t_s$, demonstrating
the entanglement of two states of the qubit with two states of the
nanomechanical resonator (see Fig.~\ref{fig5}).
\begin{figure}
\begin{minipage}{0.45\columnwidth}
\includegraphics[width=\columnwidth]{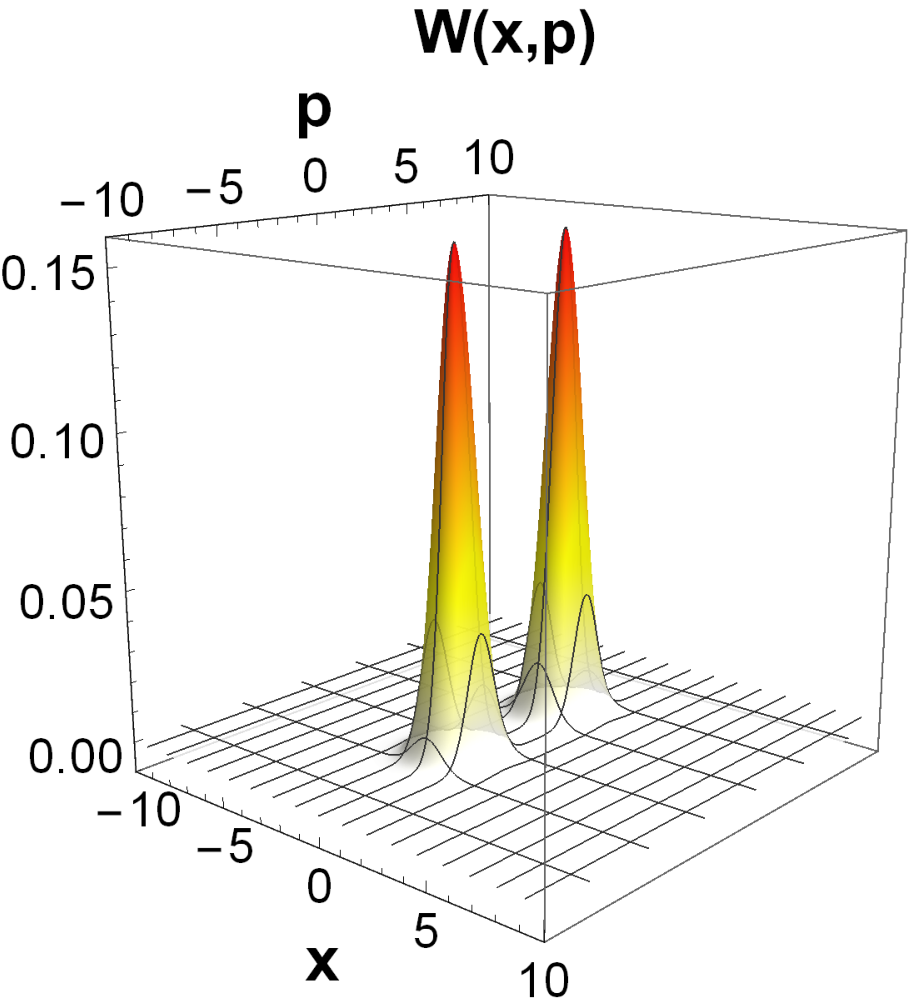}
\end{minipage}
\begin{minipage}{0.45\columnwidth}
\includegraphics[width=1.0\columnwidth]{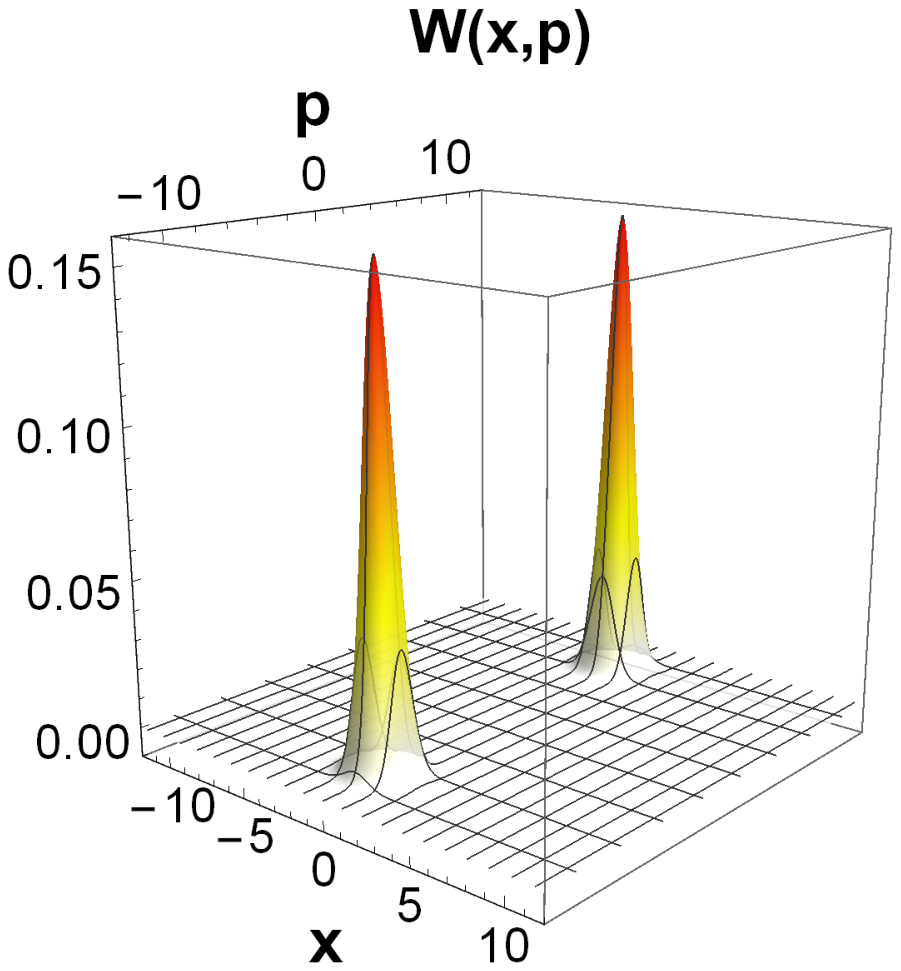}
\end{minipage}
\caption{The Wigner functions $W(x,p,t=2\pi N)$ for  $\rho=1$ (a)
and $\rho=0$ (b). It takes only positive values and have two maxima
demonstrating entanglement between two qubit states and two coherent
states of the nanomechanical resonator.}
\end{figure}\label{fig4}
\begin{figure}
\includegraphics[width=.85\columnwidth]{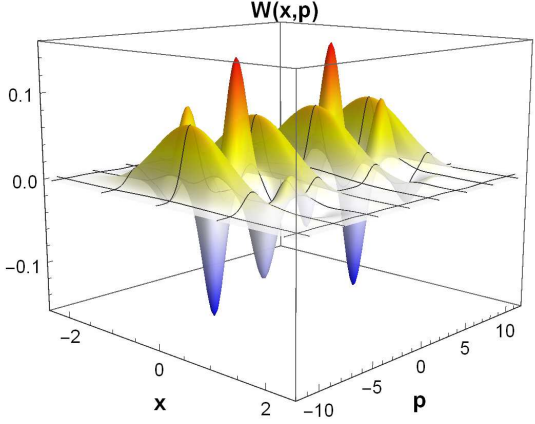}
\caption{The Wigner function $W(x,p,t=2\pi N)$ for $\rho
=1/\sqrt{2}$. It takes both positive and negative values
demonstrating entanglement between the qubit states and ``cat-states''
of the nanomechanical resonator.}\label{fig5}
\end{figure}

As it follows from the above consideration, the amplitude of
mechanical fluctuations, and therefore the energy stored in the
mechanical subsystem, changes over time.  This energy comes from the
electronic subsystem causing a rectification of ac current. To
analyze this phenomenon, we calculate the dimensionless (normalized
to $I_0=2e/\hbar$) ac Josephson current averaged over the N-th
period of the Josephson oscillations:
$$I_N=\frac{1}{2\pi}\int\limits^{2\pi N}_{2\pi(N-1)}dt
\mathrm{Tr}\left(\frac{\partial \hat H_q(t)}{\partial\Phi}\hat
\varrho(t)\right).$$  Taking into account that $\partial \hat
H_q/\partial\Phi=\eta\partial \hat H/\partial t$ and $\hat
H_q(t=2\pi N)=0$, one gets the following expression for $I_N$,
\begin{eqnarray}\label{current}
&&I_N=\frac{\eta}{2\pi}\nabla_N\mathrm {Tr}
\left(\hat H_m+\hat H_{int}\right)\hat\varrho(2\pi N) \nonumber \\
&&\hspace{0.6cm}=\frac{\eta}{2\pi}\nabla_N
\left[E_m(N)+E_{int}(N)\right],
\end{eqnarray}
where $\nabla_N f(N)\equiv f(N)-f(N-1)$ is the first difference.
From this equation, one can see that the average current is given by
the change of the mechanical energy $E_{m}$ and the energy of interaction
$E_{int}$ after N-th period. One can find that at
$N>Ns=[t_s/2\pi]+1$ the functions $E_m(N)$ and $E_{int}(N)$ can be
written as follows,
\begin{eqnarray}
&&E_m(N)=2\pi^2\tilde\varepsilon^2\left(\rho^2(2N_s-N)^2
+\tau^2 N^2\right), \nonumber \\
&&E_{int}(N)=2\pi
\varepsilon\tilde{\varepsilon}\left[\rho^2\left(N-2N_s\right)
\text{e}^{-(2\pi\tilde\varepsilon)^2(N-2 N_{s})^{2}} +\right.
\nonumber\\
&&\hspace{1.3cm}\left.+\tau^2 N\text{e}^{-(2\pi\tilde\varepsilon
N)^2 }\right].
\end{eqnarray}

The change in the interaction energy contributes to the averaged current
as well as the mechanical energy. However, this contribution is of the
order of $\tilde\varepsilon^2$ and important only for periods for
which $I(N)/\tilde\varepsilon\simeq\tilde\varepsilon^2$. So, the
average current is determined by the change of mechanical energy
mainly, and is defined  by the following equations,
\begin{eqnarray}\label{I}
&&\frac{I(N)}{\tilde\varepsilon}\approx
I_{m}(N)=-2\pi\tilde\varepsilon N,
\hspace{0.5cm} N\leq N_s-1\\
&&\frac{I(N)}{\tilde\varepsilon}\approx
2\pi\tilde\varepsilon\left( N-2\rho^2N_s\right), \hspace{0.5cm} N>
N_s.
\end{eqnarray}
\begin{figure}
\includegraphics[width=0.85\columnwidth]{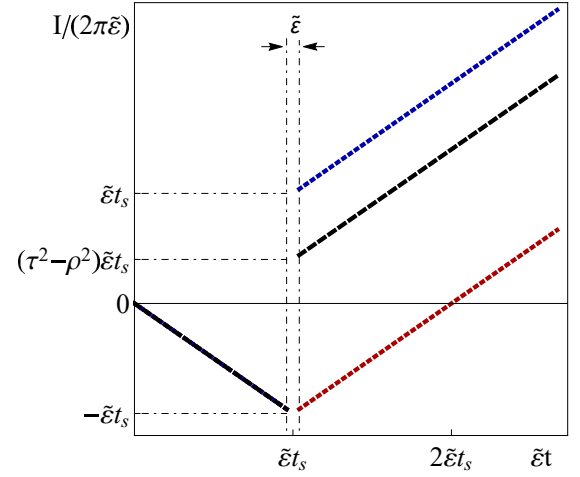}
\caption{Schematic illustration of the time-averaged Josephson
current as a function of time for different values of $\rho$ (black dashed curve). The dotted lines indicate the limiting cases of $\rho=0$ (top, blue online) and $\rho=1$ (bottom, red online) The current for $t<t_s$ does not depend on $\rho$ ($\rho=1$), see Eqs.~(\ref{I}),(\ref{Sw1}). The period, $N_s$, corresponded to the time of the bias voltage switching, is out of the consideration.}
\end{figure}\label{avc}
From Fig.~6 one can see that the averaged current exhibits a jump
equal to $-\rho^2 I(N_s)$ after the period during which the bias
voltage is switched. It originates in the fact that when we switch the
sign of the bias voltage (at $t=t_s$) the power, pumped into the
mechanical subsystem, changes depending on the magnitude of $\rho^2$.
For $\rho=1$, the supplied power, $P=I V$, just changes its sign with the bias voltage,
and the current continues to flow in the same direction as it did
before switching. For $\rho=0$ supplied power is not changed and
consequently the current direction changes after switching.

In conclusion, we have analyzed quantum dynamics of the NEMS
comprising the movable CPB qubit, subjected to an electrostatic
field and coupled to the two bulk superconductors,controlled by the bias voltage,
via tunneling processes. We demonstrate
analytically that if the ac Josephson frequency of superconductors,
controlled by the bias voltage, is in resonance with the mechanical
frequency of the CPB, the initial pure state (direct product of the
CPB state and ground state of the oscillator) evolves in time into
the coherent states of the mechanical oscillator entangled with the
qubit states. Furthermore, we established the protocol of the bias
voltage manipulation which results in the formation of entangled
states incorporating so-called cat-states (the quantum superposition
of the coherent states). The organization of such states is confirmed
by the analysis of the corresponding  Wigner function taking
negative values, while their specific features provide the
possibility for their experimental detection by measuring the
average current. The discussed phenomena may serve as a foundation
for the encoding of quantum information from charge qubits into a
superposition of the coherent mechanical states.  It may constitute
interest  for the field of quantum communications due to the
robustness of such multiphonon states regarding external
perturbation, comparing to the single-phonon Fock state. However,
the discussion of the specific protocols for such encoding is out of
the scope of this paper and will be presented elsewhere.

\section*{\emph{Acknowledgements}}
The authors thank I.V.~Krive and R.I.~Shekhter
for the useful discussions. This work was partially supported by the
Institute for Basic Science in Korea (IBS-R024-D1). LYG thanks the
IBS Center for Theoretical Physics of Complex Systems, Daejeon, Rep.
of Korea, for their hospitality.


\begin{thebibliography}{99}

\bibitem{Roukes} K.L.~Ekincia, M.L.~Roukes, {\it Review of Scientific
Instruments} {\bf 76}, 061101 (2005).
%
\bibitem{Pashkin} Y.~Nakamura, Yu.A.~Pashkin and J.S.~Tsai,
{\it Nature} {\bf 398}, 786-788 (1999).
%
\bibitem{Schmid} S.~Schmid, L.G.~Villanueva, and M.~Roukes,
{\it Fundamentals of Nanomechanical Resonators}, Springer, Switzerland
(2016).
%
\bibitem{Cleland1} K.J.~Satzinger, Y.P.~Zhong, H-S.~Chang,
G.A.~Peairs, A.~Bienfait, M.-H.~Chou, A.Y.~Cleland, C.R.~Conner,
E.~Dumur, J.~Grebel, I.~Gutierrez, B.H.~November, R.G.~Povey,
S.J.~Whiteley, D.D.~Awschalom, D.I.~Schuster, A.N.~Cleland, {\it Nature}
{\bf 563}, 661 (2019).
%
\bibitem{Cleland2} A.~Bienfait, K.J.~Satzinger, Y.P.~Zhong, H.-S.~Chang,
M.-H.~Chou, C.R.~Conner, E.~Dumur, J.~Grebel, G.A.~Peairs, R.G.~Povey,
A.N.~Cleland, {\it Science} Vol.{\bf 364}, Issue 6438, 368 (2019).
%
\bibitem{Chu} Y.~Chu, P.~Kharel, W.H.~Renninger, L.D.~Burkhart, L.~Frunzio, P.T.~Rakich, R.J.~Schoelkopf, {\it Science} {\bf 358}, 6360, Issue 199 (2017).
%
\bibitem{Cleland3} M.-H.~Chou, E.~Dumur, Y.P.~Zhong, G.A.~Peairs, A.~Bienfait, H.-S.~Chang, C.R.~Conner, J.~Grebel, R.G.~Povey, K.J.~Satzinger and A.N.~Cleland, to be published, ArXiv: 2012.04583 [quant-ph]
%
\bibitem{LaH} M.D.~LaHaye, J.~Suh, P.M.~Echternach, K.C.~Schwab and M.L.~Roukes, {\it Nature} {\bf 459}, 960 (2009).
%
\bibitem{Tian} L.~Tian, {\it Phys.Rev.B} {\bf 72}, 195411 (2005).
%
\bibitem{Gorelik2} L.Y.~Gorelik, A.~Isacsson, Y.M.~Galperin, et al.,
{\it Nature} {\bf 411}, 454 (2001)
%
\bibitem{Isacsson} A.~Isacsson, L.Y.~Gorelik, R.I.~Shekhter,
Y.M.~Galperin, and M.~Jonson, {\it Phys.Rev.Lett.} {\bf 89}, 277002 (2002).
%
\bibitem{Girvin} S.M.~Girvin,  {\it Phys.Rev.Lett.} {\bf 123}, 250501 (2019).
%
\bibitem{Girvin1} S.M.~Girvin, arXiv:1710.03179v1 [quant-ph] (2017)
%
\bibitem{NW1} K.~Masuda, S.~Moriyama, Y.~Morita,
K.~Komatsu, T.~Takagi, T.~Hashimoto, N.~Miki, T.~Tanabe, and H.~Maki, {\it Appl. Phys. Lett.} {\bf
108}, 222601 (2016).
%
\bibitem{NW2} A.~Bezryadin and P.M.~Goldbart, {\it Adv.Mater.} {\bf 22},
1111-1121 (2010).
%
\bibitem{A-A} P.~Arrangoiz-Arriola, E.A.~Wollack, Z.~Wang et al. {\it Nature} {\bf 571}, 537 (2019).
%


\end{thebibliography}
\end{document}